\documentclass[journal]{IEEEtran}
\usepackage{amsmath,amssymb,amsfonts}
\usepackage{graphicx, subfiles}
\usepackage{pgfplots}
\pgfplotsset{compat=1.18}
\usepackage[ruled,vlined]{algorithm2e}

    \usepackage[left = 1.95 cm, right = 1.95 cm, top = 1.95 cm, bottom = 1.95 cm]{geometry}

\usepackage{enumitem}    

\usepackage[english]{babel}
\usepackage[utf8]{inputenc}
 \usepackage[ruled,vlined]{algorithm2e}
 \usepackage{bm}
 \usepackage{caption} 

 \usepackage{cite}

\ifCLASSINFOpdf
\else
\fi
\usepackage{subfig}
\usepackage{amsmath,amssymb,amsfonts}

\usepackage{graphicx}
\graphicspath{%
    {converted_graphics/}
    {/}
}

  \makeatletter
\newcommand*{\rom}[1]{\expandafter\@slowromancap\romannumeral #1@}
\makeatother

\usepackage{xcolor}
  
\begin{document}
 \bstctlcite{IEEEexample:BSTcontrol}

\title{Securing the Future of IoMT in the Post-Quantum Era: An Edge-Native Federated Learning Approach\\
}

\author{Taym~Alshoghri, Deemah~H.~Tashman,~\IEEEmembership{Member, IEEE}, Mohammad Reza Gerami,~\IEEEmembership{Student Member, IEEE}, and Soumaya~Cherkaoui,~\IEEEmembership{Senior Member, IEEE}%
\thanks{T. Alshoghri was an intern in the LINCS Laboratory, Department of Computer and Software Engineering, Polytechnique Montréal, 
Montréal, QC, Canada, during this work. They are now with the Department of Computer Science, University of Toronto, Ontario, Canada, M5S 3H2 (e-mail: taym.alsho@gmail.com), D.~H.~Tashman, M. R. Gerami, and S.~Cherkaoui are with   LINCS Laboratory, 
Department of Computer and Software Engineering, Polytechnique Montréal, 
Montréal, QC, Canada, H3T 1J4 (e-mail: deemah.tashman@polymtl.ca, mohammad-reza.gerami@polymtl.ca, soumaya.cherkaoui@polymtl.ca). 
LINCS Lab website: https://lincslab.ca/en/}%
}

 \maketitle

\begin{abstract}
 Internet of Medical Things (IoMT) devices operate under strict resource constraints while handling highly sensitive health data, making security and privacy critical concerns. Federated learning (FL) further complicates this landscape, as model updates exchanged during training may unintentionally expose private medical information. Emerging quantum computing capabilities threaten the long-term viability of conventional lightweight cryptographic mechanisms, motivating the integration of Post-Quantum Cryptography (PQC) into IoMT systems. This article discusses key enabling technologies for quantum-resilient IoMT, including post-quantum key establishment, lightweight encryption, and edge-native orchestration. We propose a scalable Kubernetes-based framework that integrates PQC into FL-enabled IoMT environments and validate it on a Raspberry Pi testbed. Results demonstrate that distributed cryptographic processing significantly reduces latency compared to sequential designs while maintaining feasible resource overhead. {\color{black}The primary contribution of this work lies in the design and validation of a secure orchestration and communication framework for FL-enabled IoMT systems.}  We conclude by outlining future directions toward energy-aware architectures, intelligent security optimization, and resilient next-generation Intelligent Internet of Medical Things (IIoMT) ecosystems.
\end{abstract}


\IEEEpeerreviewmaketitle

\section{Introduction}

The Internet of Medical Things (IoMT) is transforming modern healthcare by enabling continuous patient monitoring, remote diagnostics, intelligent therapy delivery, and data-driven clinical decision making. From wearable sensors and smart implants to connected imaging systems, IoMT technologies are redefining how medical services are delivered, improving accessibility, personalization, and operational efficiency. However, this growing ecosystem introduces critical security and privacy challenges \cite{10263869}. {\color{black} Medical data is highly sensitive, while IoMT devices typically operate under strict resource constraints and are often deployed across heterogeneous and distributed environments.} These characteristics create a persistent tension between performance, scalability, and security, making traditional protection mechanisms increasingly difficult to apply. Ensuring robust, future-proof security for IoMT systems is therefore not merely a technical requirement but a fundamental prerequisite for safe and trustworthy digital healthcare. For healthcare operators, this translates into being able to scale remote monitoring, implants, and wearables without reopening security architectures every time new cryptographic threats emerge.\\
Lightweight security mechanisms are preferable for securing IoMT due to operating under stringent computational and energy constraints. However, widely deployed public-key cryptographic schemes, including RSA and Elliptic Curve Cryptography (ECC), derive their security from the computational hardness of mathematical problems such as integer factorization and the discrete logarithm problem. Moreover, these foundations are vulnerable in the presence of large-scale quantum computers, where Shor’s algorithm could efficiently break such schemes \cite{ali2025next}. Although practical quantum computers of this scale have yet to materialize, the anticipated evolution of quantum technologies raises significant concerns for long-term IoMT security. In parallel, Federated Learning (FL) has emerged as a practical solution for IoMT systems, enabling collaborative model training across distributed medical devices without centralizing sensitive patient data. {\color{black}However, although FL reduces exposure of raw data, it introduces new security risks, including privacy leakage through model updates and potential manipulation of distributed training processes \cite{zhu2019deep}.} The convergence of these challenges requires architectures that simultaneously address learning-layer vulnerabilities and future quantum threats.\\
To tackle the aforementioned challenges, a transition to Post-Quantum Cryptography (PQC) is necessary to ensure long-term security. PQC consists of cryptographic algorithms that are resistant to attacks initiated through quantum computers. {\color{black} These algorithms are based on advanced mathematical concepts, such as the Ring Learning With Errors (RLWE) problem. Unlike traditional cryptography, these mathematical problems are currently believed to be unsolvable even by powerful quantum computers.} However, PQC schemes are generally more computationally demanding than classical cryptography, a factor that is critical for resource-constrained IoMT devices \cite{8406610}. As a result, some IoMT devices may lack the hardware capabilities required to migrate to PQC, while those that can may face significant overhead. These constraints motivate the need for an implementation that minimizes overhead while providing quantum resistance.







\section{Enabling Technologies for Quantum-Resilient IoMT Systems}

{\color{black} A secure and scalable  Intelligent Internet of Medical Things (IIoMT) system depends on the coordinated integration of several complementary technologies. Cryptographic mechanisms, orchestration platforms, communication middleware, and lightweight security primitives work together to determine the system’s performance, scalability, and resilience.} This section introduces the key technologies underpinning the proposed framework, highlighting their functional roles and their relevance to resource-constrained IoMT environments. These components form the architectural backbone that enables quantum-resilient security, efficient FL coordination, and practical real-world deployment.

\subsection{Post-Quantum Key Establishment
}
Secure key establishment is a fundamental requirement in distributed IoMT systems, particularly when sensitive medical data is exchanged across heterogeneous and potentially untrusted environments. Traditional public-key mechanisms, however, are increasingly vulnerable to emerging quantum computing capabilities, motivating the adoption of post-quantum Key Encapsulation Mechanisms (KEMs). For example, Lo et al. \cite{10730752} proposed a lightweight post-quantum authentication protocol tailored for IoMT environments, demonstrating that quantum-resilient device onboarding is achievable with limited computational overhead. Such efforts highlight the feasibility of integrating PQC into IoMT communication frameworks.

Module-Lattice–based Key Encapsulation Mechanism (ML-KEM) is a NIST-standardized post-quantum key encapsulation scheme designed to securely establish shared cryptographic keys between communicating entities. {\color{black} Instead of transmitting secret material directly, ML-KEM allows two parties to establish a shared key through a secure encapsulation process.} This capability is especially important in IoMT systems, where sensitive medical data must be protected while keeping communication and computation overhead low \cite{fips2024203}. 

{\color{black}ML-KEM relies on the Modular Learning With Errors (MLWE) math problem, which is based on lattices, offering a strong mix of high security and high performance, and is designed to resist quantum computer attacks.} \cite{wang2023lattice}. For resource-constrained IoMT devices, combining ML-KEM for key establishment with a lightweight symmetric encryption scheme offers an effective approach to achieving quantum-resilient security with manageable performance overhead.

\subsection{Edge-Native Orchestration (Kubernetes)}
Managing distributed services in edge-enabled IoMT systems introduces unique orchestration challenges. Unlike cloud environments, edge platforms must operate under strict resource constraints while maintaining reliability, scalability, and fault tolerance. Container orchestration frameworks address these challenges by enabling automated workload scheduling, service recovery, and efficient resource utilization across heterogeneous devices. In this context, a container is a lightweight software package that bundles an application and its dependencies, allowing it to run consistently across different hardware platforms.

Kubernetes (K8s) has emerged as the dominant orchestration platform for containerized applications, offering capabilities such as load balancing, auto-scaling, and self-healing.  For instance, Lin et al. \cite{10592425} demonstrated how K8s-based edge orchestration can enable resource-constrained IoMT devices to participate in personalized FL, incorporating model compression techniques and validating the framework on a real-world testbed.  However, its standard distribution may impose excessive overhead for resource-limited edge deployments. K3s provides a lightweight Kubernetes distribution specifically optimized for edge and IoT environments. By retaining core orchestration features while significantly reducing memory and computational requirements, K3s becomes well-suited for Raspberry Pi-based edge clusters in healthcare scenarios. It has approximately half the memory footprint of standard Kubernetes and is performance-optimized for ARM-based architectures. A typical K3s architecture includes the following components:
\begin{itemize}
    \item \textbf{Containers:} Containers are lightweight software packages that bundle an application's code, dependencies, and runtime environment, enabling consistent execution across different hardware platforms. In Kubernetes, containers are deployed within pods.

    \item \textbf{Pods:} Pods are the smallest deployable units in Kubernetes and consist of one or more tightly coupled containers that share storage, networking, and execution context. While a pod often contains a single container, multiple containers may be grouped when close coordination is required.

    \item \textbf{Nodes:} A node is a physical or virtual machine that hosts and runs pods. Kubernetes automatically schedules pods onto available nodes based on resource availability. In our implementation, each Raspberry Pi 4 acts as a node within the cluster.

    \item \textbf{Clusters:} A cluster is a collection of interconnected nodes that collectively run containerized workloads. In Kubernetes-based systems, the cluster includes worker nodes that execute application workloads and a control plane that manages scheduling, resource allocation, and overall system state.
\end{itemize}



\subsection {Messaging and Coordination Layer}
Distributed IoMT systems rely on continuous and reliable information exchange among heterogeneous components, including edge servers, learning agents, and security services. {\color{black} Direct point-to-point communication can limit scalability, create tight coupling between components, and reduce fault tolerance.} Messaging middleware addresses these challenges by enabling asynchronous, decoupled, and resilient communication across system entities.\\
RabbitMQ is a message brokering software that handles the exchange of messages between different applications \cite{baseri2025future}. RabbitMQ can queue messages, route them to the correct recipient, acknowledge message delivery, and replicate messages across a cluster. We chose to implement our queues with RabbitMQ because of its flexibility with message routing options and reliability.

\subsection{Containerized Service Deployment} 
Deploying distributed services across heterogeneous IoMT and edge devices requires portability, reproducibility, and simplified software management. Variations in hardware architectures, operating systems, and dependencies can otherwise introduce significant integration complexity. Containerization technologies address these challenges by encapsulating applications and their execution environments into lightweight, portable units.

Docker is a widely adopted containerization platform that enables portable and reproducible application deployment. It allows services and their dependencies to be packaged into lightweight containers that can be consistently executed across heterogeneous platforms. Within the proposed framework, Docker is used to encapsulate FL and security services into containers orchestrated by Kubernetes. Docker Hub further supports efficient storage, versioning, and distribution of container images, simplifying deployment across edge-managed IoMT devices.

\subsection{Lightweight Encryption for Constrained Devices} 
Securing communications in IoMT systems requires encryption mechanisms that balance strong security guaranties with strict computational  constraints. Conventional cryptographic algorithms, while robust, may introduce excessive overhead for resource-limited medical devices. Lightweight cryptography   becomes essential for enabling practical and efficient security in constrained IoT environments. For instance, Maqsood et al. \cite{10937815} evaluated NewHope, Kyber, and XMSS on Raspberry Pi and ESP32 platforms, demonstrating that algorithm selection in IoT healthcare settings must balance encryption speed, memory consumption, and energy efficiency. Given this, El-Hadedy et al. \cite{mohamed2024securing} demonstrated the feasibility of combining post-quantum key establishment and lightweight encryption for IoMT devices. 

Ascon is a NIST-standardized family of lightweight cryptographic algorithms designed to provide authenticated encryption with minimal resource consumption. Its design emphasizes efficiency, low latency, and strong security properties, making it well-suited for IoMT deployments. The Ascon implementation adopted in this framework provides a security strength comparable to Ascon-AEAD128 while maintaining a low computational footprint.

{\color{black} In summary, these enabling technologies collectively form the architectural backbone of our quantum-resilient framework. ML-KEM ensures post-quantum key establishment, while Ascon provides lightweight encryption suited for resource-constrained IoMT devices. These security primitives are containerized using Docker and orchestrated at the edge via K3s, ensuring scalable deployment. Finally, RabbitMQ facilitates decoupled, reliable messaging between the edge infrastructure and IoMT devices, seamlessly integrating secure communication with FL workflows.}

\section{Research Gaps and Contributions}
Despite recent advances, several important challenges remain insufficiently addressed in the integration of security, scalability, and FL within IoMT environments. Prior efforts have often examined these dimensions in isolation. For example, orchestration-focused frameworks demonstrate how Kubernetes can enable scalable deployment and personalized learning at the edge, yet they typically rely on conventional cryptographic mechanisms and do not account for emerging quantum threats~\cite{10592425}. Conversely, quantum-safe FL approaches emphasize the incorporation of PQC, but their evaluations are commonly conducted on high-performance computing platforms rather than resource-constrained IoMT devices~\cite{appiah2025enhanced}. 
{\color{black}{Table~\ref{tab:comparison} summarizes the main distinctions between the proposed framework and prior studies. Unlike existing works, the proposed framework jointly integrates PQC-enabled secure FL, lightweight K3s orchestration, and real Raspberry Pi-based edge deployment within a unified IoMT architecture.}} The key contributions and insights of this work include:


\begin{table*}[t]
\caption{{\color{black}Comparison with Existing IoMT and FL Frameworks}}
\label{tab:comparison}
\centering
\renewcommand{\arraystretch}{1.2}
\setlength{\tabcolsep}{6pt}
\begin{tabular}{|c|c|c|c|c|}
\hline
\textbf{Reference} & \textbf{PQC Support} & \textbf{Edge/K3s Deployment} & \textbf{Hardware Validation} & \textbf{FL Integration} \\ 
\hline

{[8]} & No & Yes (Kubernetes-based) & Yes (real testbed) & Yes \\ 
\hline

{[11]} & Partial (Kyber + Ascon) & Yes (K3s) & Yes (Raspberry Pi cluster) & No \\ 
\hline

{[12]} & Yes (post-quantum FL) & No & No (high-performance environment) & Yes \\ 
\hline

\textbf{This Work} & \textbf{Yes (ML-KEM + Ascon)} & \textbf{Yes (K3s edge-native orchestration)} & \textbf{Yes (Raspberry Pi 4 cluster)} & \textbf{Yes} \\ 
\hline

\end{tabular}
\end{table*}

\begin{itemize}
  \item {\color{black}\textbf{Post-Quantum Secure FL Orchestration on K3s:}
We design and implement a secure orchestration and communication framework for FL-enabled IoMT systems that integrates PQC mechanisms (ML-KEM and Ascon) into a K3s-managed edge infrastructure. The proposed framework addresses the tag mismatch errors encountered in \cite{mohamed2024securing} that lead to decryption failures through improved key handling and reliable encryption workflows.}

\item \textbf{Scalable Parallel Communication with IoMT Devices:} 
Our architecture takes advantage of K3s cluster parallelism to efficiently manage secure communication with multiple IoMT devices participating in FL. Therefore, our system orchestrates both security and learning concurrently across real IoMT deployments.

\item \textbf{Hardware-Backed Performance Benchmarking:}
We present detailed benchmark results and system metrics gathered from our hybrid PQC–FL deployment on Raspberry Pi 4 and edge nodes, providing a concrete reference point for future implementations targeting quantum-resilient, edge-deployable FL systems.
\end{itemize}
\section{Quantum-Resilient Federated Learning in IoMT}

\subsection{Proposed Framework}
In the considered scenario, multiple resource-constrained IoMT devices participate in an FL process coordinated by an edge server. The edge infrastructure is responsible for aggregating model updates, managing learning workflows, and enforcing security mechanisms. Given the long-term sensitivity of medical data, the system incorporates PQC mechanisms to ensure resilience against emerging quantum-enabled adversaries. Lightweight container orchestration enables scalable deployment and coordination of learning and cryptographic services across edge nodes. Post-quantum KEMs secure device–edge communications, while lightweight authenticated encryption protects model updates and control messages. This integrated design enables secure, scalable, and practical FL operation within constrained IoMT environments.  {\color{black}This architecture is highly relevant to several practical use cases in modern healthcare settings. For example, continuous patient monitoring via bedside monitors generates high-frequency telemetry that requires real-time, secure aggregation without raw data leaving the hospital premises. Remote diagnostics using wearable sensors benefit from this system by keeping sensitive health data on the patient's device while contributing to a globally trained predictive model \cite{10622458,10592377,11370849}. Furthermore, smart implants (such as pacemakers) require extremely secure, low-latency updates. In all these scenarios, our framework ensures data privacy via FL, while safeguarding against both current and future cryptographic threats without overwhelming the devices' limited capacities.}

\begin{figure*}[!t]
   \centering
  \includegraphics[width=0.9\textwidth]{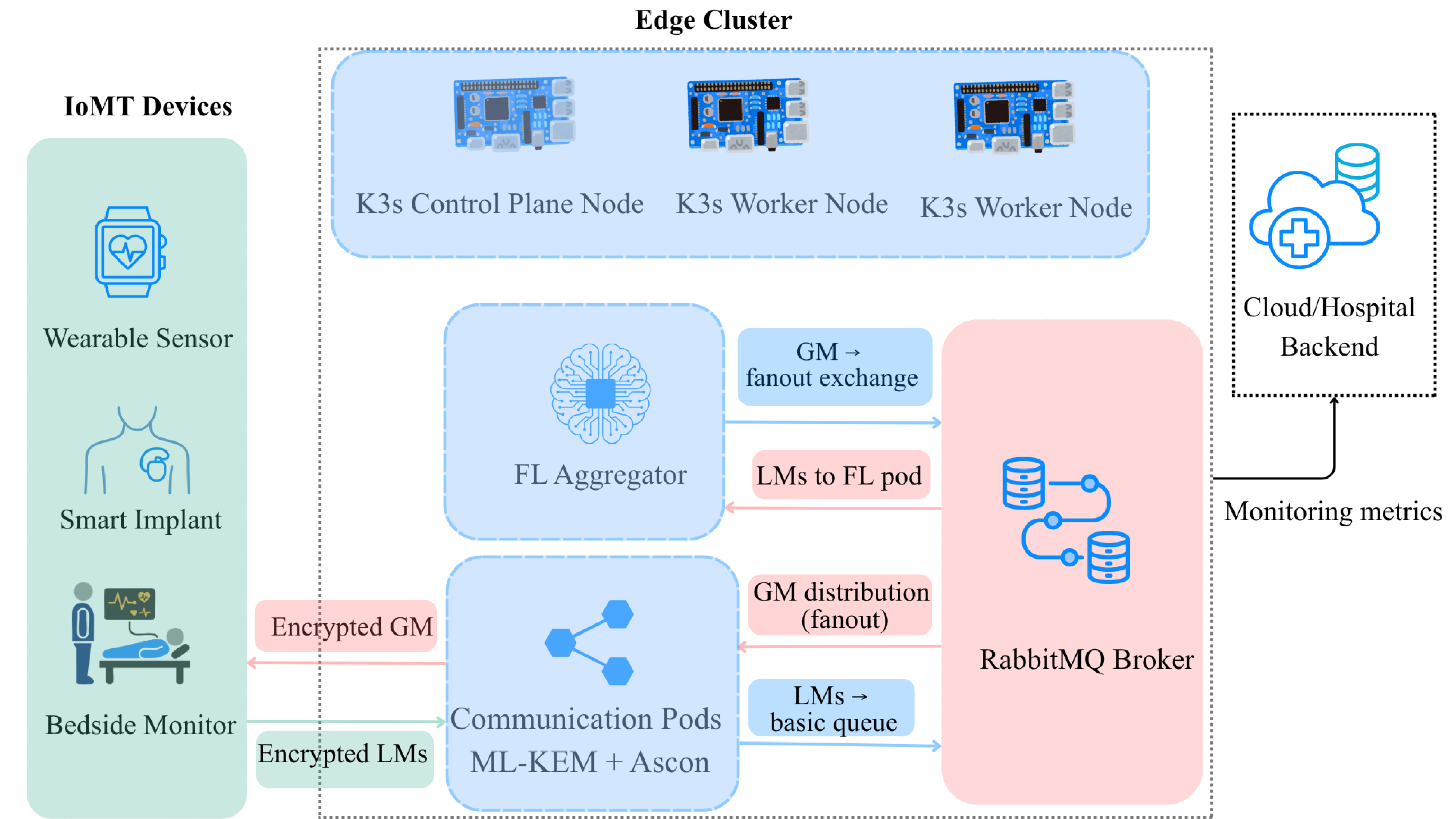}
   \caption{\color{black}{System architecture of the proposed quantum-resilient FL framework for IoMT. black blocks represent orchestration/compute components, pink blocks represent messaging/security services, and green blocks denote IoMT endpoints. The framework enables secure exchange of encrypted local models (LMs) and the global model (GM) through ML-KEM- and Ascon-enabled communication pods deployed over a K3s edge cluster.}}
   \label{arch}
\end{figure*}

Figure \ref{arch} illustrates this overall system architecture, detailing the secure data flows between the IoMT endpoints, the K3s edge cluster, and the messaging middleware.

\subsection{Experimental Testbed}
To evaluate the practical feasibility of the proposed framework, a real-world edge testbed is constructed using three Raspberry Pi 4 devices configured as a lightweight K3s cluster. Each device runs Raspberry Pi OS Lite (64-bit), a minimal operating system selected to maximize available computational resources. Within the cluster, one Raspberry Pi operates as the control plane node, while the remaining devices function as worker nodes.

The nodes are interconnected through a network switch, enabling both local cluster communication and external network access. To support interaction with services outside the cluster environment, selected applications are exposed via a NodePort service, allowing controlled external connectivity. Within the K3s cluster, we deploy three main types of pods: 
\begin{itemize}
    \item \textbf{FL Pod:} Aggregates the local models (LMs) and distribute the new global model (GM) to the communication pods. 
    
    \item \textbf{Communication Pods:}   Handle all communications with IoMT devices, including key exchange, encryption, decryption, and model transmission. One communication pod is created for each IoMT device. 
    
    \item \textbf{RabbitMQ Pod:}  Manages message queues between the FL pod and communication pods and connects IoMT devices to their assigned communication pods through the NodePort service, which exposes the cluster to the internet.
    
\end{itemize}

\noindent It is worth mentioning that the  FL pod and the communication pods are deployed using Docker Hub container images built for the ARM64 CPU, which is the CPU used in Raspberry Pi 4s. {\color{black}Although our framework does not directly implement a specific FL algorithm, it is designed to be highly flexible. By defining the aggregation and local training protocols, most standard FL schemes can be seamlessly integrated. This modular design allows researchers and developers to adapt the framework to a wide range of learning tasks and communication requirements.}  The workflow of our scenario proceeds as follows:



\begin{enumerate}
 
    \item Each communication pod establishes a shared key with its assigned IoMT device using ML-KEM.

    \item The FL pod distributes the GM to the communication pods.

    \item Using the shared key, each communication pod encrypts the GM and sends it to its IoMT device.

    \item Each IoMT device decrypts the GM, performs local training on its private data, and encrypts and updates the model before transmission.

    \item Communication pods decrypt the received LMs, update and forward them to the FL pod.

    \item The FL pod aggregates the collected local updates to produce a new GM.

    \item Steps 2--6 are repeated for the desired number of training rounds.
\end{enumerate}
\begin{figure} [b]
    \centering
   \includegraphics[width=0.5\linewidth]{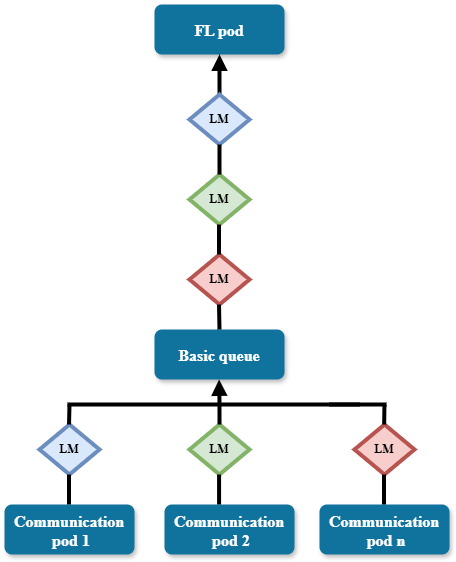}
    \caption{{\color{black} Basic queue design for Local Models (LMs) aggregation. Each communication pod publishes its decrypted LM into a shared basic queue, from which the central FL pod consumes the models for global aggregation.}}
    \label{basic}
\end{figure}

\begin{figure}
    \centering
\includegraphics[width=0.5\linewidth]{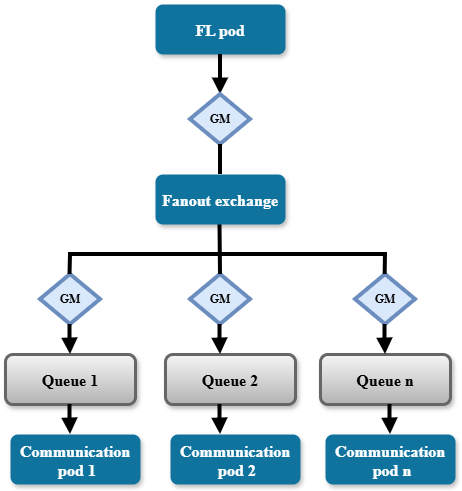}
    \caption{{\color{black}Fanout exchange design for Global Model (GM) distribution. The central FL pod publishes the newly aggregated GM to a fanout exchange, which duplicates and routes the model to all subscribed communication pods simultaneously.}}
    \label{fanout}
\end{figure}

\begin{figure}
    \centering    \includegraphics[width=0.5\linewidth]{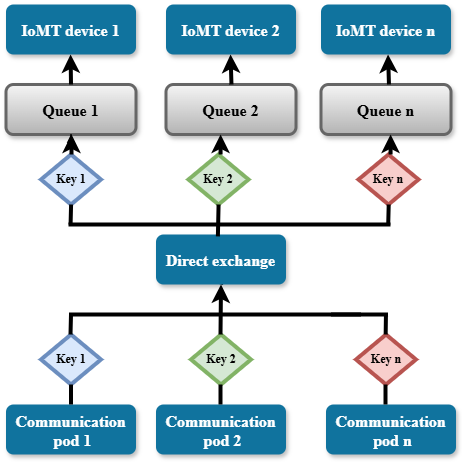}
    \caption{{\color{black}Direct exchange design for secure communication between communication pods and assigned IoMT devices. This setup ensures isolated, one-to-one communication channels utilizing unique routing keys for each device.}}
    \label{direct}
\end{figure}
\subsection{RabbitMQ}
We use RabbitMQ to facilitate communication between the communication pods, FL pod, and IoMT devices through the use of queues and exchanges. Within our architecture, we define three types of RabbitMQ exchanges:
\begin{itemize}
    \item \textbf{Basic Queue:}  Used to allow the communication pods to send their decrypted {\color{black}LMs} to the FL pod. Each communication pod publishes the LMs into the basic queue while the FL pod consumes from the queue until enough models are received for aggregation  (Figure \ref{basic}).

    \item \textbf{Fanout Exchange:}   Used to distribute a new GM to the communication pods. First, each communication pod subscribes a queue to the fanout exchange. When the FL pod sends a GM to the fanout exchange, RabbitMQ copies the message and publishes it to each subscribed queue. As a result, each communication pod receives the updated GM near simultaneously  (Figure \ref{fanout}). 

    \item \textbf{Direct Exchange:} The direct exchange mechanism is responsible for secure one-to-one communication between each communication pod and its assigned IoMT device. To ensure that messages are delivered exclusively to the intended recipient and can be correctly decrypted using the established shared key, each pod–device pair is assigned a unique routing identifier during initialization. Queues subscribing to the direct exchange bind using this identifier as the routing key. When a message is published, it includes the corresponding routing key as metadata, enabling RabbitMQ to forward the message only to the queue with a matching binding. This design guarantees isolated and secure communication channels between each communication pod and its paired IoMT device (Figure \ref{direct}).
\end{itemize}

RabbitMQ queues are potentially a single point of failure in this architecture, which could raise security concerns should the queues fail or become temporarily unavailable. To account for this, we use quorum queues wherever possible. Quorum queues are durable, replicated queues that are optimized for data safety and high availability. The contents in a quorum queue are maintained after restarts and are fault-tolerant through use of the Raft consensus algorithm.

\subsection{Post-Quantum Cryptography Configuration}
{\color{black}We adopt the ML-KEM-512 parameter set instead of ML-KEM-768 or ML-KEM-1024 to better suit the resource constraints of IoMT environments. ML-KEM-512 and Ascon both provide approximately 128-bit classical security. Therefore, selecting higher ML-KEM security parameter sets would increase computational and communication overhead without proportionally improving the overall security level of the proposed framework, since the end-to-end protection remains aligned with Ascon’s security strength.  
Consequently, ML-KEM-512 represents a practical tradeoff between quantum-resistant security and lightweight performance, enabling resource-constrained IoMT devices with limited computation and energy capabilities to participate efficiently in the framework. Regardless of the selected parameter set, the proposed distributed architecture still provides significant acceleration compared to conventional sequential approaches.}

\section{Results \& Analysis}

 In this section, we provide the results   of our proposed scenario. To evaluate the efficiency of the proposed distributed architecture, its performance is compared with a \textit{standard sequential} approach. In the sequential design, a single program executes all operations—including key exchange, encryption, decryption, and model distribution—in a strictly consecutive manner for each IoMT device. 
In contrast, the proposed \textit{distributed} architecture assigns these tasks to multiple communication pods, enabling parallel execution of key exchange and cryptographic operations across devices. {\color{black}The sequential baseline was selected to isolate and quantify the impact of distributed orchestration and parallel cryptographic execution within the proposed framework. While several distributed FL and edge orchestration frameworks exist in the literature, relatively few studies investigate the integration of PQC mechanisms within lightweight K3s-based IoMT edge deployments combined with practical Raspberry Pi testbed validation. Therefore, the sequential implementation provides a controlled reference point for evaluating the latency reduction and scalability benefits introduced by the proposed distributed architecture.}

\begin{figure*}[t]
    \centering
    {
          \includegraphics[width=0.48\textwidth]{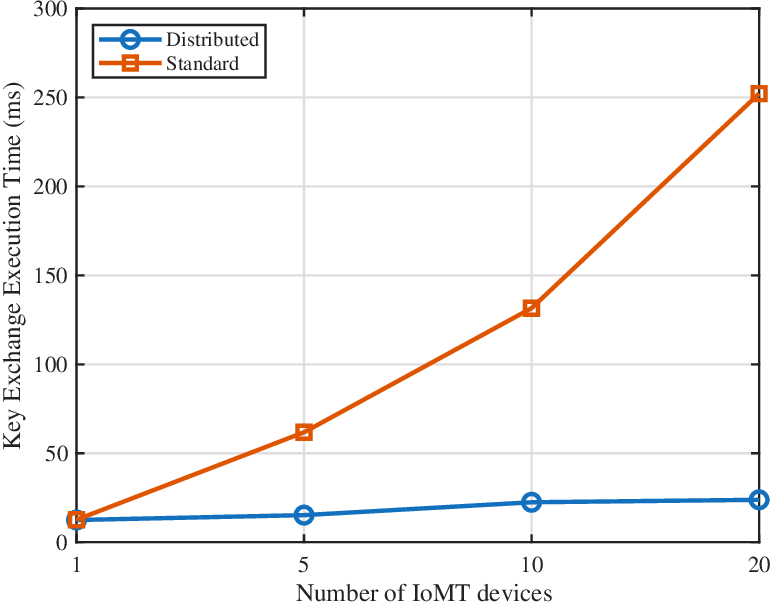}
        \label{fig1a}
    }
    \hfill
    {%
        \includegraphics[width=0.48\textwidth]{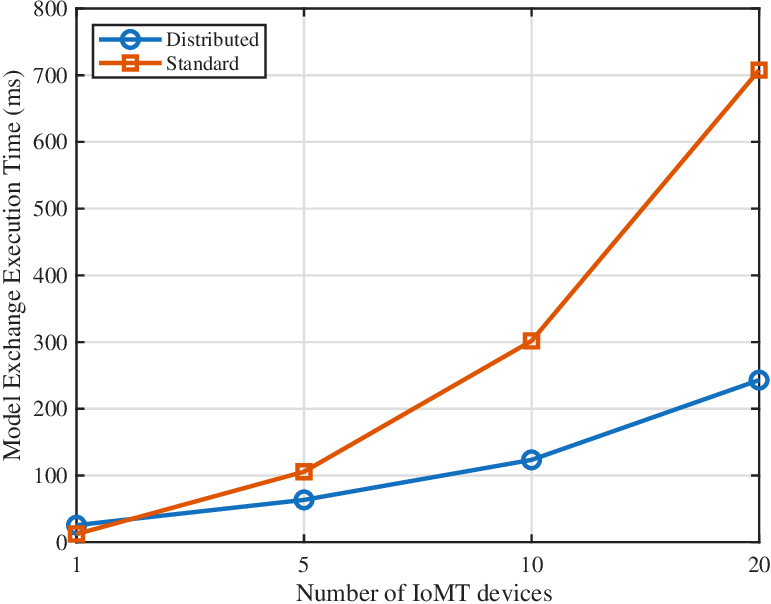}
        \label{fig1b}
    }
    \caption{{\color{black}Execution time (ms) versus the number of connected IoMT devices for (a) key exchange and (b) model exchange. The proposed distributed architecture maintains lower, near-constant latency through parallel processing compared to the linearly increasing latency of the standard sequential approach.}}
    \label{fig:combined}
\end{figure*}

\par Figure \ref{fig:combined} (a) reports the total key exchange latency observed for all IoMT devices, defined as the elapsed time between the initiation of the exchange process and the completion of the final device interaction. 
The results highlight a clear difference in scalability behavior. While the standard sequential approach exhibits an execution time that increases with the number of IoMT devices, the distributed framework maintains an approximately constant latency. This contrast underscores the effectiveness of workload distribution and parallel processing in mitigating device-scale performance degradation. This behavior is particularly beneficial for IoMT security, where maintaining predictable cryptographic latency is essential for ensuring timely protection of sensitive medical data.


Figure \ref{fig:combined} (b) presents the execution time for model exchange as the number of IoMT devices increases. For the distributed architecture, latency is measured from GM distribution by the FL pod to the reception of local updates, while for the standard scheme it spans GM encryption to LM decryption. Both approaches experience increasing execution time as more devices participate; however, the standard scheme grows sharply, whereas the distributed architecture increases more gradually, demonstrating improved scalability through parallel processing. 

The results reveal a clear scalability advantage for the proposed distributed framework. When more than five IoMT devices are present, significant reductions in latency are observed across both key exchange and model exchange processes. For example, with ten devices, latency reductions of approximately 83\% and 59\% are achieved for key exchange and model exchange, respectively. With twenty devices, reductions of approximately 91\% and 66\% are observed. At lower device counts, coordination overhead becomes more pronounced. While performance gains remain observable at five devices, their magnitude is reduced. In the single-device scenario, the distributed architecture exhibits higher latency due to queue-based coordination overhead. As system scale increases, however, the parallel distribution of cryptographic workloads outweighs these overhead effects, resulting in substantial system-level acceleration.

{\color{black}It is worth mentioning that  the integration of post-quantum cryptographic primitives within edge-assisted IoMT frameworks may increase computational and communication overhead, particularly for battery-powered medical sensors and wearable devices. Future work will therefore investigate empirical power-consumption profiling and energy-aware optimization strategies using the Raspberry Pi-based edge testbed to better characterize the tradeoff between security resilience, latency, and energy efficiency in practical deployments.}

 \section{Emerging Directions in IoMT}

\subsection{Energy-Aware Quantum-Resilient Architectures in IoMT}
Post-quantum security mechanisms, while essential for future-proof IoMT systems, may introduce additional energy demands that challenge the sustainability of constrained medical devices. Designing energy-aware quantum-resilient architectures therefore emerges as a critical research direction. Future systems may incorporate adaptive security scheduling, lightweight cryptographic parameter tuning, or intelligent computation offloading to maintain IoMT devices longevity without compromising protection. 

\subsection{Joint Optimization of Learning and Security for IoMT}
The coexistence of FL and PQC presents new opportunities for intelligent system optimization. Learning-enabled IoMT devices equipped with deep reinforcement learning (DRL)-based decision engines could dynamically balance model training, communication overhead, and cryptographic processing. Such adaptive mechanisms may enable IoMT systems to respond autonomously to variations in workload, channel conditions, and energy availability while preserving quantum-resilient security guarantees. This convergence of AI-driven control and post-quantum security represents a promising frontier for intelligent medical IoT systems.

\subsection{Trustworthy and Attack-Resilient IoMT Systems}
While PQC mechanisms enhance confidentiality and key establishment, future IoMT systems must also address vulnerabilities arising at the learning and system levels. FL environments remain susceptible to adversarial behaviors, including model poisoning, malicious updates, and compromised devices. Future research may therefore explore the integration of quantum-resilient cryptography with adversarial AI defenses and anomaly detection mechanisms.

\subsection{Interoperability and Standardization for Quantum-Resilient IoMT}

As post-quantum mechanisms transition from research to deployment, interoperability and standardization will become critical concerns. IoMT ecosystems consist of heterogeneous devices, vendors, and communication protocols, requiring harmonized security frameworks that support seamless integration. Future work may investigate standardized quantum-resilient communication stacks, lightweight cryptographic profiles, and deployment guidelines tailored for constrained healthcare environments. Addressing interoperability challenges is essential for enabling scalable, vendor-agnostic quantum-safe IoMT infrastructures.

{\color{black}\subsection{Limitations and Deployment Challenges}
While the proposed framework significantly enhances security and scalability, it presents certain deployment challenges. First, the centralized messaging broker (RabbitMQ) could act as a single point of failure; although mitigated through fault-tolerant quorum queues, broker outages could temporarily disrupt FL rounds. Second, implementing PQC on severely resource-constrained medical devices introduces unavoidable energy overhead, which may reduce the battery life of continuous monitoring wearables or implants. Finally, the heterogeneous nature of healthcare ecosystems poses interoperability challenges, requiring standardized cryptographic profiles to seamlessly integrate various vendor-specific IoMT devices into a unified K3s cluster.}

\section{Conclusions}
In this article, we first discussed the growing need for PQC in IoMT environments, where sensitive healthcare data and long device lifecycles make future quantum threats particularly concerning. Moreover, we outlined the key enabling technologies that support quantum-resilient deployments, including post-quantum key establishment, lightweight encryption, and edge-native orchestration. Building on these foundations, we proposed a Kubernetes-based FL framework that integrates scalable cryptographic processing with containerized service deployment and coordinated messaging. 
Experimental results demonstrated that the distributed architecture significantly reduces key exchange and model exchange latency compared to standard sequential designs, while maintaining scalability as the number of IoMT devices increases. Finally, we highlighted future research directions toward energy-aware designs, intelligent optimization of learning and security, attack resilience, deployment challenges, and interoperability for next-generation quantum-resilient IoMT ecosystems.


 

\end{document}